# Green Computer Science Millennial Students Examination

Leon Andretti Abdillah
Department of Information Systems, Universitas Bina Darma
Jalan Ahmad Yani No. 3
Palembang 30264 INDONESIA

*Abstract*- **Smartphones as one of information technology products have been affected higher education in various aspects. This article explains the useness of smartphones in facilitating online examination in information systems and computer science students. The research objective to be achieved by the researchers through the research, are as follows: 1) Utilizing smartphone as a media test online exam for green computing environment, 2) How to use social information technologies in online test, and 3) Explore the facilities or features that could be used for the online exam implementation. The observation was conducted with 100 students as respondents. Researcher used google forms to disseminate questions for online examination. The findings of the research showed that most the college students used Android OS for their online examination. Social technology like google forms has rich features in supporting online examination for computer science students. The use of smartphones, google forms, and facebook can create an atmosphere of modern, green computer science exams, efficient, and environmentally friendly.**

## I. INTRODUCTION

Information technology (IT) offers many wonderful applications over internet. IT provides new space for digital world activities. Statistic shows that digital population worldwide as of July 2018 [1] has passed 4 billion users. Figure 1 informs us that: 1) Active internet users are 4.119 billion, 2) Unique mobile internet users are 3.802 billion people, 3) Active social media users are 3.35 billion, and 4) Active mobile social media users are 3.096 billion users.

Current information technology has been collaborated with many sectors, such as business, government, health care, entertainment, politics, including IT-based education. Group of social technology or social media has been attracted author's attention for a long time [2]. Several researchers have studied social media in politics [3]-[4], business [5]-[6], and personality [7]-[8]. This article extends the previous article that cover online examination by using smartphones [9] and social technology [10]. IT linkaged blended learning environment through social technology, pervasive mobile device and cloud technology [11].

Recent IT development have been merged with communication and education sectors. In communication field, IT becomes main backbone for serving the movement of data among gadgets. In education field, IT serves the learning process between lecturers and students. Since mobile communication device used in the middle of 1990s, many aspects in our daily life change including transportation [12]-[15], travelling [16]-[18] and residential locations [19]-[21].

Smartphones are being adopted at a phenomenal pace [22] because smartphones have a diverse set of media capture capabilities [23]. Smartphones have many advantages in sending voice, text, image, data, in rinch format and extreme speed. As handheld personal computers, smartphones represent the most recent step in the evolution of portable information and communication technology [24]. Smartphone devices can be used both as a mobile telephone and as a handheld computer [25], that containing a wide range of sensors and communications interface [26].

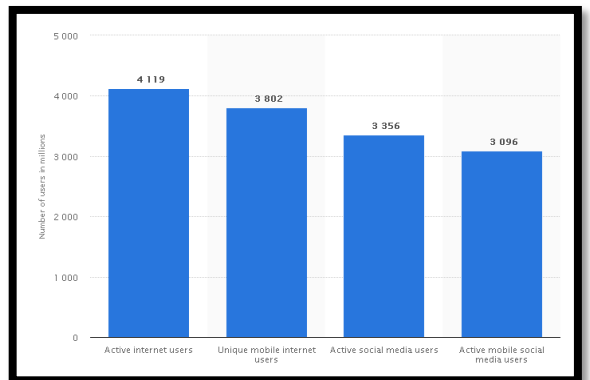

Fig. 1. Social media usage per July 2018.

This ubiquitous technology is viewed as a versatile device [27] that dominantly used by young people or millennial generation, a cohort we define here as adults ages 18 to 29 [28]. The Millennials were born in or after the year 1982 [29] until 2002 [30]. This generation also known by Baby Boomer and Gen X generations [31]. The term of "Millennials" is the winner based on the online polling by Peter Jennings from ABC News Tonight [32]. Nielson reported that people own smartphone per age group [33] could be seen in table 1.

Social media users are led by surged by more than 20 percent in 2018 [34], with Facebook in particular posting impressive increases, despite already being the world's most popular social platform for the past decade. Now, close to 2.8 billion people around the world use social media at least once a month, with more than 91 percent of them doing so via mobile devices.

TABLE I
TYPE SIZES FOR PAPERS

| No | Age | Percentage |
|---|---|---|
| 1 | 18-24 | 98% |
| 2 | 25-34 | 97% |
| 3 | 35-44 | 96% |
| 4 | 45-54 | 89% |
| 5 | 55-64 | 80% |
| 6 | 65+ | 68% |

The prediction by January 2018 [34]: 1) Total number of active social media users are 2.789 billion or eual to 37% of total population, and 2) Total number of social users accessing via mobile are 2.549 billion or equal to 34% of total population (see Figure 2).

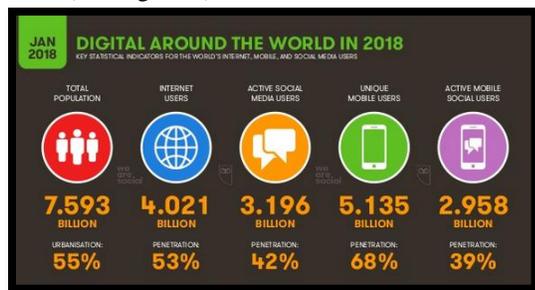

Fig. 2. Social media usage per January 2018

Among various of social media, market leader Facebook keep stayed on top of the rank [35] followed by YouTube, WhatsApp, Facebook Messenger, etc. Facebook was the first social network to surpass 1 (one) billion registered accounts and currently sits at 2.05 billion monthly active users (Figure 3).

Social media users are led by surged by more than 20 percent in 2018 [34], with Facebook in particular posting impressive increases, despite already being the world's most popular social platform for the past decade. Last but not least, smartphones can act as green technologies and as integral parts of green information systems [23].

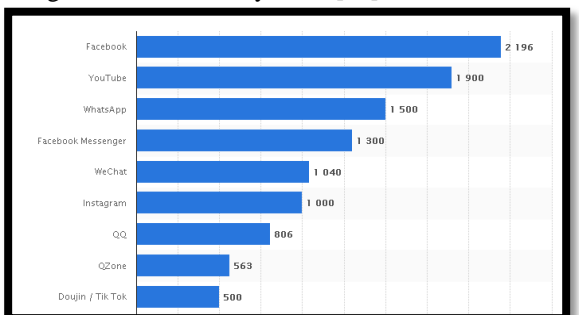

Fig. 3. Leading social media per August 2018.

Such as a personal computer (PC) or laptop, smartphones also operated by an operating systems (OS). The OS is the heart of a smartphone software system [36]. Today Smartphones OS are dominated by five major players, namely: 1) Android OS - Google Inc., 2) iOS - Apple Inc., 3) Seri 40 OS - Nokia Inc., 4) BlackBerry OS - BlackBerry Ltd., and 5) Windows OS - Microsoft Corporation. According to International Data Corporation (IDC) [37], in the 2016 Q2, Android lead the market shares until 87.6% followed by iOS, Windows, etc.

The rest of this article is organized into 3 (three) more sections. The next section, Section 2, will discuss research methods. After that, Section 3, covers results and discussions. This article is closed with a conclusion as Section 4.

## II. RESEARCH METHODS

### A. Respondents

Total respondents involved in this research amounted to 100 freshmen (early year students) who are taking courses of Knowledge Management Systems. This course is the main course on one of his concentrations in information systems (computer science faculty) fields, Knowledge Management Systems.

### B. Action Observation

Author implements the observation in the class to analyze the benefits of smartphones in mobile students examination. Student seating positions are arranged according to their serial number in the attendance list. During the exam did not use paper at all because the test is done by using facilities provided by google form. The exam media involves a smartphone.

### C. Exam Questions

The examinations are given 3 (three) times during a semester: 1) Daily Examination, 2) Middle Examination, and 3) Final Examination. All of those examinations consist of 10 (ten) questions.

The questions posed is currently online exam consists of three types, namely: 1) Multiple choices questions (four choices form a to d), 2) Dichotomy questions (provide only two possible answers, True or False), and 3) short answer questions (type less than three words answers).

The examination need to be done in 1 (one) hour and 15 (fifteen) minutes. The first 10 (ten) minutes are used to fill students information. The next 1 (one) hour is allocated for answering all of the 10 (ten) questions. And the last 5 (five) mintes are used for final check and submitting the answers.

### D. Online Google Form

This study involves several application of social information technology, such as: 1) Google Forms as the medium for online examination, 2) FaceBook, 3) WordPress, and 4) DropBox.

Google Form is used as a medium by lecturers to provide examination questions. Google forms could be used to record and display students' responses [38]. This application normally for online surveys [39], but later on could be used for many purposes or "one size fits all" [40].

Lecturer creates a questions sheet by using google forms. After all questions with the possible answers' space ready, then lecturer send virtual form via 1) email, 2) Uniform Resource Locator (URL) or shorten URL, 3) Embed HTML. Lecture also able to share the online questions forms to 1) Google+, 2) Facebook, or, 3) Twitter. Lecturer needs to do setting for collecting students responses via google forms. Lecturer needs to set or make the google forms as "quizzes". In quiz options, lecturer able to release the mark immediately or later after manual review (turns on email collection). Respondent or students can see: 1) Missed questions, 2) Correct answers, and 3) Point values.

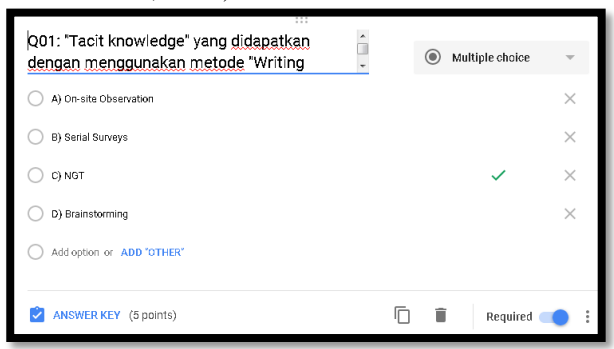

Fig. 4. Answer key and points.

Students will answers the questions through the given link via social media facebook. Faceboook is used to create social media-based learning environment, group students-based on their class, disseminate serial lecture contents, announce a number of information, provided examination URL.

Students will answer all of the questions from their mobile phone, personal computer (PC), or laptop. They need to input their valid google email into the google forms. Fill all the required fields in google forms. After students finished submitting their answers, a confirmation response will display a confirmation message. Each question in this study has equal mark, 10 points. Google forms able to save the correct answer for automatically marking (see Figure 4).

## III. RESULTS AND DISCUSSIONS

According to the observations in the class, researcher able to synthesize some point of views as results. In the first result author would like to display a post in facebook. This post is used to inform about the exam to the students. The lecturer needs to provide the date of the test, especially the time limit.

Lecturers also need to provide a URL that will take students to google forms containing exam questions online. Current facebook able to display the preview from te URL (Figure 5). After student clicks the given URL then browser will take it to google forms containing examination virtual sheet.

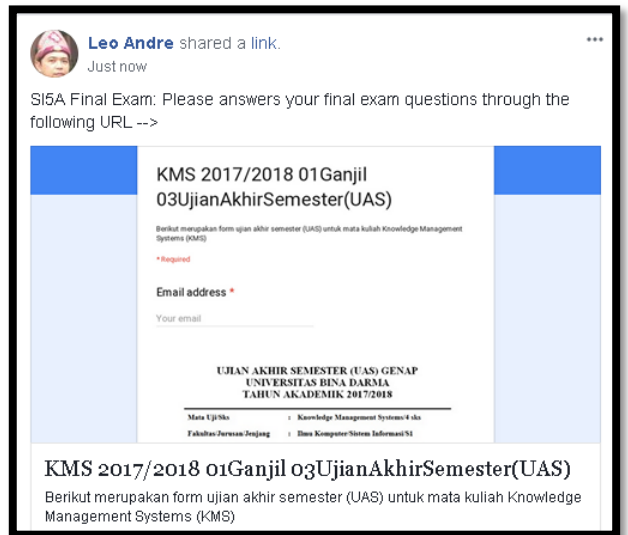

Fig. 5. Facebook post of online examination.

### A. Students' Characteristics

The number of students as respondents in this study amounted to 87 early year computer science students. Students who were respondents in this research are dominated by male students (53.9%), while female students amounted to 46.1%.

TABLE II
RESPONDENT INFORMATION

| Criteria | Sub Criteria | Percentage |
|---|---|---|
| Gender | Men | 53.9% |
|  | Female | 46.1% |
| Status | Students | 89.7% |
|  | Private company | 4.6% |
|  | Civil servants | 2.3% |
|  | Self-employed | 1.1% |
| Gadgets | Smartphone Android | 54.0% |
|  | PC|Laptop | 40.2% |
|  | Smartphone Windows | 4.6% |
|  | Smartphone iPhone | 1.1% |
|  | Smartphone Android | 54.0% |

Although most of the respondents were students (83.9%), but some of them have worked. Approximately 4.7% worked in private companies, 2.3% work as civil servants (PNS) or state-owned enterprises (BUMN), 1.2% self-employed, while the other 2.3%.

### B. Students Identification Sheet

In virtual google forms there are a number of fields that must be filled out by the students. For this study, all the fields are mandatory. When student arrive into google form for their examination, they will found notes that consist of class code, exam type, exam instruction, time limit, and the URL.

Fig. 6. Students' Identification Form.

The first 3 (three) fields are mandatory in: 1) Email address, 2) Student registration number or NIM, and 3) Student full name. The sign of asterisk (*) means that the field is mandatory.

*C. Online Questions Sheet*

The first type of questions is questions that require short answers. Lecturer limits the answer will be given by the students as much as a maximum of two words. Place to answer short questions provided under the given problem. Information scale scores for each question appears on the upper right. In this reserach, each question is 10 points.

Fig. 7. Multiple choice questions example.

The second type of questions is dichotomy. This type of questions has only two possible answers (true or false), see Figure 7.b for te example of this questions. Students need to click one of the best answer for this type of question. Type of dichotomy questions is basically a type of multiple-choice questions that has only two possible choices of answers.

The third type of questions are multiple choice questions. In this type, there are four possible answers her. Each possible answer is coded in the form of the character 'a', 'b', 'c' or 'd'. (Figure 7). At the end of the google virtual form there is "Submit" button which enables students to submit all of the answers.

*D. Respondent Focus*

Based on observations at the time of the mobile based exam, students focus on their own smartphone when answering the given questions. Because the space in smartphones are relative narrow than personal computer or laptop, students have less pedagogical movement to change active screen. They will focus on signing their google account, complete personel identification, then answering 16 questions, folowed by some evaluation questionnaires.

*E. Real Time Response*

After students finished answering their answers and then click submit from their active google form, the information will be appeared in admin/lecturer screen. Lecturer able to see who is the latest student that submit is/her answers through google form. In this research, author uses google forms to record and display students' responses. The following figure shows how google forms display recently submitted answers from a student in real time.

*F. Colorful Bar Chart Representation*

One of google form facility is statistic visualization in bar chart or pie diagram. The statistical view of bar chart is enriched with percentage and legend of each part of the chart. The section of correct section of the chart will also signed by different colours (darker).

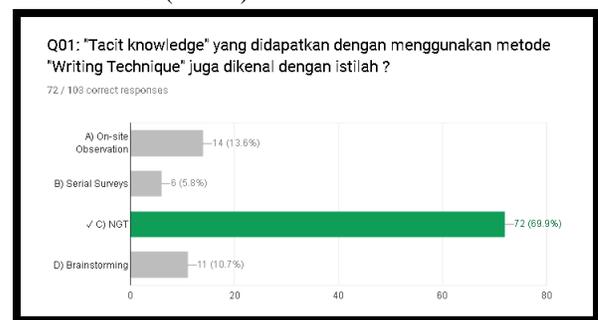

Fig. 8. Multiple choice questions example.

Figure 8 shows an example of visual representation by using tab chart enriched with percentage. Figure 8 informs 69.9% of students anwers are correct. The correct choice will be displayed in a darker color.

*G. Robust Tabulation*

After all students finished submitting their answers, google forms provide the facilities to open the respons in spreadsheet (create a new spreadsheet or select existing spreadsheet). The spreadsheet is similar to microsoft excel spreadsheet. Once we have the data in Excel then we are able to do some tabulations.

Fig. 9. Authomatic tabulation.

### H. Online Examination Evaluation

At the end of te exam, students must submit their respon to the exam scheme. The first 3 (three) questions are related to comfortable aspect for student. According to their responses, majority of students are "very comfortable" and "comfortable" with online exam, smartphone-based exams.

TABLE III
RESPONDENT COMFORTABILIT

| Criteria | VC | C | QC | QC | Uc |
|---|---|---|---|---|---|
| comfortable with the online exam | 45.3% | 46.5% | 8.1% | - | - |
| comfortable with smart-phone-based exams | 33.7% | 50.0% | 12.8% | 3.5% | - |
| comfortable with the smartphone-based online exam | 30.2% | 53.5% | 16.3% | - | - |

The author also collects opinions from students about the types of questions asked during the online exam. The percentage type of questions most liked by students are "Multiple choices" of 42.7%. Meanwhile, the questions type that students do not like is "None" for 24.3%.

TABLE IV
FAMOUS QUESTIONS TYPE

| Percentage type of questions most liked | The type of questions | Percentage type of questions least liked |
|---|---|---|
| 26.2% | All | 15.5% |
| 20.4% | Short Questions | 19.4% |
| 6.9% | Dichotomy | 19.4% |
| **42.7%** | Multiple choices | 21.4% |
| 3.9% | None | **24.3%** |

The percentage for learning and examination venue could be seen in table 5. The students who "Likes" and "Really like" to study in the class are 68.9% and 22.3%. Meanwhile for study in laboratory, 49.5% of students are "Likes" and 35.9% are "Really like" (please see Table 5).

TABLE V
STUDY IN THE LABORATORY

| The percentage of students enjoys studying in the classroom | The type of study venue | The percentage of students enjoys studying in the laboratory |
|---|---|---|
| 22.3% | Really like | 8.7% |
| **68.9%** | Likes | **49.5%** |
| 6.8% | Quite like | 35.9% |
| 0% | Do not like it much | 1.9% |
| 1.9% | Do not like | 3.9% |

The percentage for examination location could be seen in table 6. The students who "Likes" and "Really like" to have exam in the class are 67.4% and 18.6%. Meanwhile for having exam in laboratorium, 59.3% of students are "Likes" and 24.4% are "Really like" (please see Table 6).

TABLE VI
EXAM IN THE LABORATORY

| The percentage of students love the exam in the classroom | The type of exam venue | The percentage of students love the exam in the laboratory |
|---|---|---|
| 18.4% | Really like | 7.8% |
| **67%** | Likes | **52.4%** |
| 11.7% | Quite like | 32% |
| 1% | Do not like it much | 4.9% |
| 1.9% | Do not like | 2.9% |

The last evaluation is related to students comfortability in online smartphone-based examination by using google form. The percentage of "Comfortable" equal to 55.3% followed by "Very comfortable" for 30.1%, and some 12.6% are "Quite Comfortable" to use smartphone in their examination (see Figure 10).

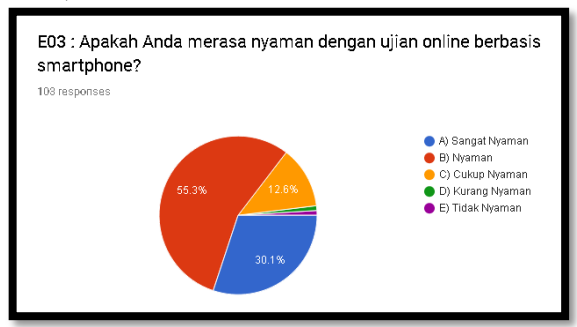

Fig. 10. Examination by using google form evaluation.

## IV. CCONCLUSION

Based on the observation reported above, author reach to several following conclussions, as follow: 1) Smarphone-based examination become new trend in IT-based education activities, 2) Mobile examination has personal impact in promoting green base education, 3) Mobile examination trigger students to concentrate more on their own question, 4) Mobile examination robust correcting process, 5) Dominat students are comfortable with smartphone-based exams, 6) Most of the students love all types of the questions esspsecially "Short Questions", 7) Most of the students loves to study and having exams both in the class and/or in the laboratory, 8) Most of the students also enjoys to study by involving social media facebook, and last but bot least 9) This evaluation still has some limitations.